\documentstyle[12pt]{article}
\begin{document}
\newpage
\begin{center}
\Large {EFFECT OF DISSIPATIVE FORCES ON THE THEORY OF A SINGLE-ATOM MICROLASER}\\
\end{center}
\vskip 0.8cm
\begin{center}
\large{
N. Nayak} \\ 
\end{center}
\begin{center}
\large{
S. N. Bose National Centre for Basic Sciences,}\\ 
\end{center}
\begin{center}
\large{
Block-JD, Sector-3, Salt Lake City, Calcutta-700091, India }\\
\end{center}
\vskip 0.5in
\begin{center}
\bf {Abstract}\\
\end{center}
\indent
\baselineskip=24pt
We describe a one-atom microlaser involving Poissonian input of atoms
with a fixed flight time through an optical resonator. The influence
of the cavity reservoir during the interactions of successive individual
atoms with the cavity field is included in the analysis. The atomic decay
is also considered as it is nonnegligible in the optical regime. During the
random intervals of absence of any atom in the cavity, the field evolves
under its own dynamics. We discuss the steady-state characteristics of 
 the cavity field. Away from laser threshold, the field can be
nonclassical in nature. 
\vskip 0.5in
\newpage   
\noindent
The subject of one-atom laser, optical counterpart of the micromaser, has
generated extensive interest after the recent experimental demonstration
by An et al$^1$. In the experiment, two-level $^{138}Ba$ atoms in their upper
states are pumped into an optical cavity in such a way that average number
of atoms in the resonating mode satisfies the condition $<N>\leq 1.0$. The average number
of photons $<n>$ in the mode shows a linear dependence on the pump until
$<N>\simeq 0.6$. But, a further increase in $<N>$ displays a thresholdlike jump
in $<n>$. Calculations based on a one-atom theory, such as the one given in
Ref.1, can explain the linear regime only.
An and Feld$^2$ incororated the cavity mode structure into the one-atom theory
for an explanation of this jump. Kolobov and Haake$^3$ addressed to the problem
by a Poissonian pumping model having partial overlaps of travel times of
successive atoms in the cavity. Alternative explanations$^{4,5}$ have
also been proposed to explain the thresholdlike structure.\\
\indent
In this paper, we address to a slightly different pump mechanism in the
microlaser setup: atoms are streamed into the cavity in such a way that
strictly one atom can pass through it at a time. The pumping is Poissonian
with the intervals between successive atomic flights being random.
Thus, the flight time through the cavity is fixed for each atom. In
other words, what we are considering here is an exact optical counterpart
of micromaser setup$^6$. The one-atom laser theories available in literature,
for example in Ref. 7, analyze steady-state properties of the Jaynes-Cummings
interaction$^8$ in a damped cavity$^9$. The one-atom theory by Fillipowicz
et al$^{10}$ needs recasting as it does not consider atomic damping which is
important in the optical regime and also due to the cavity enhancement
factor, the socalled Purcell factor. In addition, the theory disregards 
leakage of radiation from the cavity during the atom-field interaction. The
magnitudes of cavity dissipation constant and atom-field coupling
constant in the microlaser experiment$^1$ indicate that radiation
dissipation from the optical resonator is nonnegligible during the
entire dynamics irrespective of presence of any atom in the cavity.\\
\indent
The above factors lead us to the need of a theory capable of handling
reservoir-induced interactions and the Jaynes-Cummings interactions$^8$
simultaneously. The theory proposed by the present author in Ref. 11
may be suiatable for the present purpose.The theory assumes the
atom-field coupling constant independent of cavity mode structure
which is alright for the microwave cavity$^6$. In fact, such a situation
is necessary for the generation of nonclassical fields$^4$. In the
optical cavity in Ref. 1, atoms travel through a length of about
40 wavelenghts of the interacting mode and thus making the coupling
constant dependent on the mode structure. However, a novel and simple technique
in streaming the atoms through the cavity, adopted by An et al$^{12}$
to improve their earlier setup$^1$, provides an uniform atom-field
coupling constant in the cavity. This setup with single-atom events
would be suitable for generating nonclassical fields in the optical
regime. With such a system in mind, we follow the method in Ref. 11.
\\
\indent 
We assume atoms arrive individually at the cavity with an average
interval $\bar{t}_{c}=1/R$ where $R$ is the flux rate of atoms. We have
$t_{c}=\tau + t_{cav}$ where $\tau$ is the interaction time, fixed
for every atom, and $t_{cav}$ is the random time lapse between one
atom leaving and successive atom entering the cavity. $\bar{t}_{c}$
is the average of $t_c$ taken over a Poissonian distribution in time
 of incoming atoms. The cavity field evolves by this repititive
 dynamics from near vacuum as thermal photons in the optical cavity
 are almost nonexitent.
  Thus, during $\tau$ , we have to solve the equation
 of motion
 \begin{eqnarray}
\dot{\rho} &=& -i[H,\rho ]-\kappa (a^{\dagger}a\rho -2a\rho a^{\dagger}+
\rho a^{\dagger}a)
\nonumber \\
& & -\gamma (S^+ S^- \rho -2S^- \rho S^+ +\rho S^+ S^-)
\end{eqnarray}
where $H$ is the Jaynes-Cummings Hamiltonian$^8$ and $\kappa$ and $\gamma$
are the cavity-mode and atomic decay constants respectively. $a$
is the photon annihilation operator and $S^+$ and $S^-$ are the Pauli
pseudo-spin operators for the two-level atomic system. During $t_{cav}$, the cavity field evolves under its own dynamics represented by Eq. (1) with
$H=\gamma =0$. The method for obtaining
a coarse-grained time derivative, valid for a Poissonian process$^{13,14}$,
for the photon number distribution $P_n =<n\vert \rho \vert n>$ is given
in detail in Ref. 11. The steady-state photon stattistics is then
\begin{equation}
P_{n}=P_{0}\prod_{m=1}^{n}v_{m}
\end{equation}
and $P_{0}$ is obtained from the normalisation $\sum^{\infty}_{n=0}P_{n}=1$.
The $v_{n}$ is given by the continued fractions
\begin{equation}
v_n=f^{(n)}_{3}/(f^{(n)}_{2} + f^{(n)}_{1}v_{n+1})
\end{equation}
with $f^{(n)}_{1}=(Z_{n}+A_{n+1})/\kappa$, 
$f^{(n)}_{2}=-2N+(Y_{n}-A_{n})/\kappa$ 
and $f^{(n)}_{3}=-X_{n}/\kappa$.
 $N=R/2\kappa$ is the number of atoms passing through the cavity in a 
photon lifetime.
$A_{n}=2n\kappa$ and
 $X_{n}$, $Y_{n}$ and $Z_{n}$
are given by
$$X_{n}=R\sin^{2}(g\sqrt{n}\tau)\exp\{-[\gamma+(2n-1)\kappa]\tau\},$$
\begin{eqnarray}
Y_{n} &=& \frac{1}{2}R\Bigl(\{2\cos^2[g\sqrt{n+1}\tau]-\frac{1}{2}
(\gamma/\kappa+2n+1)+F_{1}(n-1)\}\exp\{-[\gamma
\nonumber \\
& & +(2n+1)\kappa]\tau\}+
[\frac{1}{2}(\gamma/\kappa+2n+1)-F_{2}(n-1)]\exp\{-[\gamma+
(2n-1)\kappa]\tau\}\Bigr),
\nonumber
\end{eqnarray}
and
\begin{eqnarray}
Z_{n} &=& \frac{1}{2}R\Bigl([\frac{1}{2}(\gamma/\kappa+2n+3)+F_{2}(n)]
\exp\{-[\gamma+(2n+1)\kappa]\tau\} 
\nonumber \\
& & -[\frac{1}{2}(\gamma/\kappa+2n+3)+F_{1}(n)]\exp\{-[\gamma
+(2n+3)\kappa]\tau\}\Bigr).
\nonumber
\end{eqnarray}
The functions $F_{1}$ and $F_{2}$ are:
\begin{eqnarray}
F_{i}(n)&=&\frac{\kappa/4g}{(\sqrt{n+2}-\sqrt{n+1})^2}\Biggl[\frac{\gamma}{\kappa}(\sqrt{n+2}-\sqrt{n+1})\sin{(2g\sqrt{m}\tau)}
\nonumber \\
& & -\frac{\gamma}{g}\cos{(2g\sqrt{m}\tau)}-[2n+3+2\sqrt{(n+1)(n+2)}](\sqrt{n+2}-\sqrt{n+1})\sin{(2g\sqrt{m}\tau)}\Biggr]
\nonumber \\
& & +\frac{\kappa/4g}{(\sqrt{n+2}+\sqrt{n+1})^{2}}\Biggl[\pm\frac{\gamma}{\kappa}(\sqrt{n+2}+\sqrt{n+1})\sin{(2g\sqrt{m}\tau)}
\nonumber \\
& & -\frac{\gamma}{g}\cos{(2g\sqrt{m}\tau)}\mp[2n+3-2\sqrt{(n+1)(n+2)}](\sqrt{n+2}+\sqrt{n+1})\sin{(2g\sqrt{m}\tau)}\Biggr]
\nonumber
\end{eqnarray}
where $m=n+2$ and $n+1$ for $i=1$ and $2$ respectively with the upper sign for $i=1$.
Once $P_n$ is obtained, we can describe the characteristics of the cavity field
by evaluating its various moments.\\
\indent
We find that the photon statistics of the cavity field given by Eqs. (2) and
(3) is a function of the dimensionless parameters $N$, $\kappa /g$ and
$\gamma /g$. The theory in Ref. 10 does not take into account the atomic relaxation. As the cavity dissipation during $\tau$ is neglected, the photon distribution function derived there is dependent on $\kappa$ only through the parameter $N=R/2\kappa $. Hence, in the context of microlaser, it is difficult to make a proper judgement of the dissipative effects on the photon statistics from their work$^{10}$. Ref. 11 discusses in detail the degree of influence of the reservoir-induced interactions on the steady-state photon statistics. The study indicates that it is nonnegligible for the optical cavities of the type used in Ref. 1 and 12. In fact, we find that the photon statistics using results from Ref. 10 (broken-dotted curves in Figs. 1 and 2) mostly differ from the present results. The pump parameter $D=\sqrt{N}g\tau $ is introduced here as we find it
 useful for the description of microlaser characteristics.
 The structures in $<n>$ as $D$ is varied for fixed $N$ in Fig. 1
reflect the characteristics of the Jaynes-Cummings interaction$^8$. Soon after
the threshold is attained at about $D=1.0$, the photon number rises sharply.
The reason is as follows. The field is almost in vacuum before the very first
atom enters the cavity. Thus the atoms initially in their upper states contribute
varying fractions of their energies to the cavity and at $n=N$ and $D=\pi /2$ the atoms
find themselves completely in their respective lower states. Thus $<n>$ is peaked at about $D=1.6$ depending on
$\kappa$, $\gamma$ and $N$. For higher $\kappa$ and $\gamma$, the peak moves
slightly towards higher $D$ due to threshold being attained at higher $D$.
This happens even for increasing $R$ which is due to increase in the percentage of time $R\tau\times 100\% $ in the duration of one second taken by $R$ atoms  
in interacting with the cavity field resulting in the increase in dissipation
of energy to the atomic reservoir. we further find that $<n>=0$ near
 $D=$31.4, 62.8, $94.2,....$ giving $g\tau =\pi$, $2\pi$, $3\pi$,....
 respectively. At such values of $g\tau $, the atom absorbs the photon
 it has emitted before leaving the cavity.\\
 \indent
 Fig. 2 shows that the variance of the cavity field v$=\sqrt{[(<n^2>-<n>^2)/<n>]}$
 increases sharply at about $D=1.6$ where $<n>$ is also peaked [Fig. 1]. We
 find that near this value of $D$, the $P_n$ is doubly peaked at $n=0$ and
 $n\simeq N$, examplified in Fig. 3, and this increases the variance in
 photon number. However, for slightly higher value of $D$, the cavity
 field is highly sub-Poissonian in nature [Figs. 2 and 3]. It also appears 
  for further higher values of $D$. The sub-Poissonian nature of the cavity field appears because the flight time $\tau$ is short compared to the atomic lifetime $(2\gamma)^{-1}$ as well as the photon lifetime $(2\kappa)^{-1}$, discussed in detail in Ref. 11. Such situations, depicted in Fig. 2, donot happen in conventional lasers. Thus, we find that the photon field characteristics in the present microlaser case [Figs. 2 and 3] are different from that for a conventional laser [Ref. 15]. Further, it may be noticed in Fig. 1 that $<n>$ gets very small as $D$ is increased which is due the fact that the increase in interaction time $\tau$ increases the influence of atomic as well as cavity reservoirs on the atom-field interaction. The sub-Poisssonian nature of the cavity field appears for such values of $D$ since the situation $\tau <(2\tau)^{-1}, (2\kappa)^{-1}$ is met in the range of $D$ depicted in Figs. 1 and 2.
\\
 \indent
 The cavity field intensity, proportional to $<n>$, saturates 
  as $N$ is increased for fixed $\tau$. The saturation value of $<n>$ depends on
  $\tau $ as dictated by the Jaynes-Cummings
 interactions$^8$. This saturation character
 is not seen in Fig. 1 where the increase in $D$ is due to increase in
 $\tau $  for fixed $N$.\\
 \indent
 We have analysed the characteristics of a one-atom microlaser capable
 of generating nonclassical optical fields in the realistic regimes of
 atomic and cavity dissipations (e.g. $\kappa /g=0.01$ and $\gamma /g=0.1$). We have also displayed results for values of $\kappa$ such as $\kappa /g=0.001$ and $0.0001$ in the Figs. 1-3 in order to show its influence on the photon statistics which can be attained if the cavity Q-factor in the present experimental setup is enhanced by a factor of 100 or more. In addition, we have included the influences
 of atomic as well as field reservoirs on the coherent atom-field interaction.
  Also, we have considered the case of
 a Poissonian pumping of atoms into the cavity which is the case in
 the experiments. However, the parameter $N$ in the present study
 has to be distinguished from $<N>$ in those experiments$^{1,12}$
 where the statistical averaging is taken over $N_0 $ =1, 2, 3, .....
 atomic events. The arrival of $N_0 $ atoms at the cavity is, however,
 Poissonian. Yang and An$^4$ in their attempt to analyse the
 experimental results$^1$ have considered upto $N_0 = 15$. Whereas,
 in the present paper $N$ is just a simple addition of single atom
 events in the photon lifetime $(2\kappa )^{-1}$. The technique
 adopted in Ref. 12 is already capable of selecting atoms having
 a particular velocity. Further refinement should be possible to
 restrict the dynamics to only $N_0 =1$ events. Then it would be
 possible to have nonclassical optical fields of the type in Fig. 2.\\
\indent
The author would like to thank Professor G. S. Agarwal for suggesting this
problem.
 
\newpage
{\bf Figure Captions:}  \\
Figure 1: Cavity field intensity, proportional to $<n>$, as a function of the pump parameter $D$ for $N=100$ and $\gamma /g=0.1$. $\kappa /g=0.01$ (full line), $0.001$ (broken line), and $0.0001$ (dotted line). The broken-dotted line represents the results from Ref. 10 in which $\kappa =\gamma =0.0$ during $\tau$ .\\
Figure 2: v versus $D$. The parameters are same as in Fig. 1. For clarity, the broken-dotted curve is shifted upwards by 3.0. The horizontal lines are lines v$=1.0$. The sub-Poissonian nature of the radiation field is indicated by v$<1.0$.\\
Figure 3: Photon distribution function $P(n)=P_n$ for $N=100$, $\gamma /g=0.1$ and $\kappa /g=0.001$. $D=1.7$ (full line) and $D=10.0$ (broken line).\\
\end{document}